\documentclass[final,5p,times,twocolumn]{elsarticle}

\usepackage[amssymb]{SIunits}
\usepackage{color}
\usepackage{lscape}
\usepackage[normalem]{ulem}
\usepackage{amsmath}
\usepackage{makecell}

\def\tsc#1{\csdef{#1}{\textsc{\lowercase{#1}}\xspace}}
\tsc{WGM}
\tsc{QE}

\begin{document}
\let\WriteBookmarks\relax
\def\floatpagepagefraction{1}
\def\textpagefraction{.001}






\title{X-ray measurements of gas distribution in a zero gap alkaline water electrolyzer}


\author[inst1,inst2]{On-Yu Dung}

\affiliation[inst1]{organization={Power and Flow group, Department of Mechanical Engineering, Eindhoven University of Technology},
            addressline={PO Box 513}, 
            city={Eindhoven},
            postcode={5600 MB}, 
            country={the Netherlands}}
            
\author[inst3]{Stephan Boden}
\author[inst1,inst2,inst5]{Albertus W. Vreman,\corref{cor1}}
\ead{a.w.vreman@tue.nl}
\cortext[cor1]{Corresponding author}
\author[inst1,inst2]{Niels G. Deen}
\author[inst3,inst4]{Markus Schubert}
\author[inst1,inst2]{Yali Tang}

\affiliation[inst2]{organization={Eindhoven Institute for Renewable Energy Systems (EIRES), Eindhoven University of Technology},
            addressline={PO Box 513}, 
            city={Eindhoven},
            postcode={5600 MB}, 
            country={the Netherlands}}

\affiliation[inst5]{organization={Nobian Industrial Chemicals},
            addressline={P. O. Box 75730}, 
            city={Amersfoort},
            postcode={3811 LP}, 
            country={the Netherlands}}

\affiliation[inst3]{organization={Institute of Fluid Dynamics, Helmholtz-Zentrum Dresden-Rossendorf},
            city={Dresden},
            postcode={01328}, 
            country={Germany}}

\affiliation[inst4]{organization={Chair of Chemical Process Engineering, Technische Universität Dresden},
            city={Dresden},
            postcode={01062}, 
            country={Germany}}

\begin{abstract}
X-ray radioscopy was used to measure the 2D projected dynamic void fraction in a zero/narrow gap alkaline water electrolyzer at a spatial resolution of 15 $\mu$m, for narrow gap sizes up to 300 $\mu$m and current densities up to 0.54 A/cm$^2$. As expected, the void fraction in the bulk was found to increase along the cell height and with increasing current density. The void fraction measured in the gap region (the space between the diaphragm and the electrode and its holes) was always larger than in the bulk. It hardly depended on the gap size at current densities below 0.3 A/cm$^2$. The lowest cell potential was measured for zero gap. No evidence of isolating gas pockets/films in the gaps was found. Liquid crossover and oxygen void fraction exceeding the hydrogen void fraction occurred for porous plate electrodes, but these phenomena were suppressed for perforated foil electrodes.

\end{abstract}








\maketitle


\vspace{-0.2cm}
\section{Introduction}
\label{sec:intro}
Water electrolysis to generate hydrogen with renewable energy is a key technology towards a carbon-free society. Among different types of technology, alkaline water electrolysis is cost-effective and of reasonably good efficiency (see e.g. \cite{IEA2019}). The efficiency can be further improved if ohmic losses are reduced. Zero gap between electrodes and the diaphragm is considered to minimize ohmic losses (e.g. see \cite{Phillips2016}). However, the ohmic/area resistance in zero gap cells is often found much higher than expected, i.e. the much higher than the diaphragm/separator resistance \cite{deGrootVreman2021}. 

It is well-known that the hydrodynamics of the electrogenerated bubbles inside the cell can affect the cell potential in nonzero gap (flat plate) electrolyzers \cite{hreiz2015electrogenerated}. In water electrolysis, dissolved hydrogen and oxygen gases are produced separately at the cathode and anode respectively. Upon supersaturation of the dissolved gases, H$_2$ and O$_2$ bubbles are formed \cite{hreiz2015electrogenerated}. Bubbles between electrodes can act as electric insulators amd lengthen the path needed for the ions to travel from one electrode to another \cite{Tjaden2016}, thus increasing the cell resistance while electrode coverage by bubbles can reduce the active surface area of the electrode \cite{hreiz2015electrogenerated}. On the other hand, as mentioned in Ref. \cite{zarghami2020cfd}, when bubbles detach from the electrode surfaces, a microconvection at the boundary layer is induced, enhancing the mass transfer and thus also the cell efficiency \cite{vogt2011gasTheory,vogt2011gasNumerics}.

In particular, for the zero-gap (or narrow gap) configuration, the origin of the high area resistance was speculated to be caused by bubbles or a gas film trapped in the gap as practically it is almost impossible to reach a true zero gap (see also \cite{tobias1959effect,kienzlen1994location,nagai2003existence,lavorante2015water}). It is expected that the void fraction at the gap increases with decreasing gap size. Therefore there might exist an optimal distance between the electrodes for optimal cell efficiency while the area resistance contributed by the solution is still significantly reduced \cite{tobias1959effect}. The existence of the optimal gap size is experimentally found in Ref. \cite{nagai2003existence,Haverkort2021}. Furthermore, recently, experiments were reported by Haverkort \& Rajaei \cite{Haverkort2021}, in which a gap width of $\unit{200}{\mu m}$ led to a substantially lower cell voltage than a gap width of zero, suggesting that introducing a deliberate gap allows bubbles or gas films to escape more easily. Nevertheless, the presence of gas bubbles/films in `zero' gap remains still a hypothesis to date. To clear up the speculations in literature and the mystery of bubble/gas effects on the cell resistance in zero/narrow gap configurations, it is of primary interest to visualize and measure the gas distribution.


Measuring the gas distribution in electrolyzers is challenging due to opaque dense bubbly flows and the required high spatial resolution.
As mentioned in Ref. \cite{hreiz2015electrogenerated}, the attempts to measure the void fraction in which the bubbles produced by vertical plane electrodes are via image analysis (e.g. Ref. \cite{Abdelouahed2014,Rox2023} employed high speed camera) and local intrusive probes to measure the resistivity of the bubbly electrolyte solution \cite{Riegel1998,Abdelouahed2014}. The image analysis method can yield significant error near the electrode due to bubble shadowing effect and the resistivity method assumes homogeneity of void fraction in depth \cite{hreiz2015electrogenerated}. In short, measuring the void fraction in narrow-gap cells poses practical challenges due to limited optical access and spatial resolution. In this work, we employ 2D X-ray radioscopy to measure the void fraction in an electrolyzer, including the gap region with a resolution down to $\unit{15}{\mu m}$. In the past, X-ray radioscopy and tomography have been employed to study bubbly flows in porous transport layers of polymer electrolyte water electrolyzer (summarized in \cite{Yuan2023}) and proton-exchange membrane water electrolysis (summarized in Ref. \cite{Jeon2023}). X-ray measurements have also been extensively employed to dense bubbly flows in many other engineering applications \cite{Mandalahalli2020}. To the best of our knowledge, no attempt before the current work has utilized X-ray scanning to measure the gas volume fraction in zero/narrow gap electrolyzers.  

Another phenomenon that will be addressed in this paper is the liquid crossover through the diaphragm. Liquid crossover is a cause of dissolved gas crossover. Gas crossover has become an ongoing industrial challenge due to the resulted risk in safety and the gas purity. Yet very limited research on this phenomenon is available in open literature \cite{Haug2017,Hu2022,Barros2024crossover}. One transport mechanism is through the electro-osmotic flow induced by the electric potential difference across the diaphragm, which is observed in Ref. \cite{Haverkort2020PRA,Haverkort2020JPowerAdv}. In this work, we try to determine and understand the crossover in a zero-gap cell using electrodes with different porosity and surface wettability. 

The experimental setup and method is elaborated in Sec. \ref{sec:setup}. Then, the overall structure of the bubbly flows and the typical current-voltage curve are illustrated in Sec. \ref{sec:overall_structure}. After that, the effect of gap size and current densities on the measured void fraction are uncovered quantitatively in Sec. \ref{sec:gap_size_effect_to_bubble}. This is followed by a further analysis on the gas contribution to the cell voltage in Sec. \ref{sec:gapSize_bub_effect_Ecell}. In Sec. \ref{sec:cross_flow} the crossover phenomenon is addressed by comparing the different hydrodynamic behaviors in zero-gap configurations using two different electrodes.


\vspace{-0.2cm}

\section{Experimental Setup and methods}
\label{sec:setup}

\subsection{In-house built cell}
A cell was built that fitted into the X-ray scanner. Fig. \ref{fig:setup}a shows the details of the cell design. The electrodes are nickel plates with thickness $\unit{0.5}{mm}$ unless otherwise specified. The active dimensions of each electrode are $\unit{9.4}{cm}$ in height $\times$ $\unit{0.96}{cm}$ in width, which give the apparent area of $\unit{9.0}{cm^2}$. It has a porosity of 0.5 (see Fig. \ref{fig:foil_plate_drawing_configuration}a for the technical drawing). We employed Agfa's Zirfon PERL UTP 500 diaphragm ($\unit{500}{\mu m}$ thick) \cite{agfa2020technical}. To create a gap between the separator and the electrodes, we placed spacers (PEEK) in between with each being \unit{50}{\mu m} or \unit{100}{\mu m} thick, which allows for adjustable gap sizes between 0 to \unit{300}{\mu m}. The spacers can be placed either between the electrodes and the diaphragm or outside the electrodes (see Fig. \ref{fig:setup}b).

Lastly, throughout this study, the electrolyte used is potassium hydroxide (KOH) solution with an initial concentration of 24 wt\% KOH and an initial temperature around room temperature before starting a series of electrolysis experiments (elaborated below). The experiments were performed under ambient pressure.

\subsection{X-ray radioscopy measurement and the cell voltage measurement}
\begin{figure*}
    \centering

    \includegraphics[width=1.3\columnwidth]{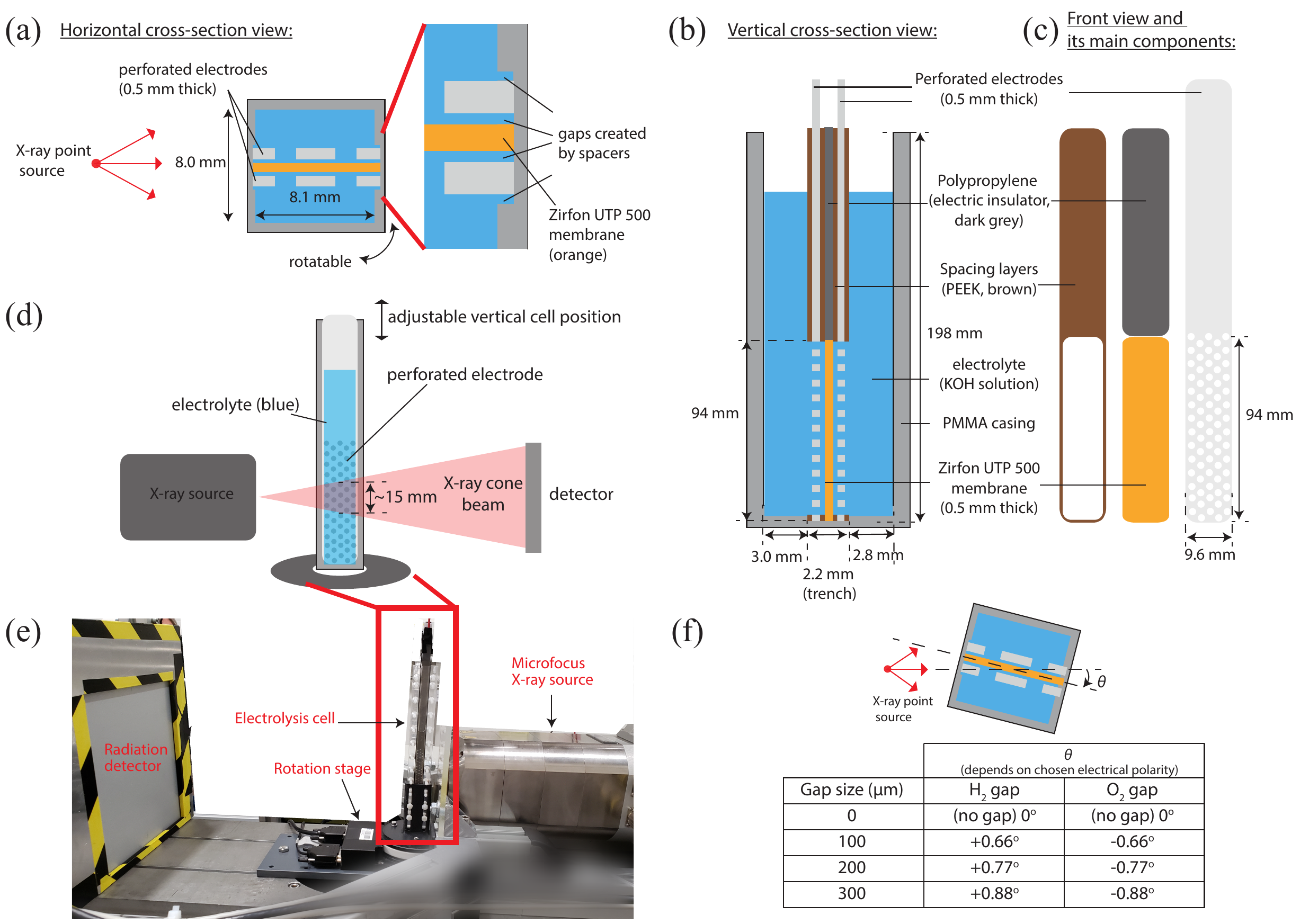}
    \caption{(a) Horizontal cross-section, (b) vertical cross-section and (c) front view of the in-house built electrolyzer including the compositions and the corresponding dimensions. (d) Schematics of the setup for the X-ray radioscopy. (e) The setup for X-ray radioscopy. (f) The cell rotation angle (anti-clockwise positive) for different gap sizes and gaps (H$_2$ or O$_2$). The angle was chosen such that the vector parallel to the membrane center (inclined dashed line) but through the center of the gap pointed towards the X-ray source point.}
    \label{fig:setup}
\end{figure*}

We employed X-ray radioscopy to measure the void fraction in the in-house built electrolyzer. Fig. \ref{fig:setup}d shows the details of the X-ray setup in this work and Fig. \ref{fig:setup}e shows a picture of the setup in operation. For the X-ray radioscopy system, we employed a two-dimensional flat panel X-ray image detector (PerkinElmer XRD 0822 AP3 IND; 205 $\times$ 205 mm$^2$ total active area covered by 1024 $\times$ 1024 pixels) and a microfocus X-ray tube (XrayWORX XWT - 190 - TC) \cite{Porombka2021}. 

The X-ray source is practically a point source with a focal spot size of less than \unit{5}{\mu m} and generates an X-ray cone beam. The distance between the X-ray source and the centre of the electrolyzer was \unit{30}{mm}. The distance to the X-ray detector was choosen to be \unit{400}{mm} such that each detector pixel covers an beam cross-section of roughly 15 $\times$ 15 µm² in the center of the electrolyzer. Thus, the detector covers roughly \unit{15}{mm} of the height of the electrolyzer. The exposure time of the detector was set to \unit{200}{ms} at the beginning of each radioscopic scan and was set to \unit{600}{ms} further on. The frame rate is the inverse of the exposure time, i.e. \unit{5}{fps} and \unit{1.67}{fps} respectively.


To create an electric potential difference between the electrodes, they are connected to a direct current power supply (KORAD, KA3005P) with a setup accuracy of $\leq 0.1\% + \unit{3}{mA}$ current operated in constant current mode. For the cell voltage measurement, we measure the voltage across the electrodes by a National Instrument data acquisition module (NI-9215, 16 bit, $\pm \unit{10}{V}$ range with 0.02 percentage error) that converts the analogue voltage signal to a digital signal. The acquisition rate is 100 Hz.

\subsection{Experimental procedure and image analysis} \label{sec:void_frac_and_exp_procedure}

\begin{figure*}
    \centering
    \includegraphics[width=1.3\columnwidth]{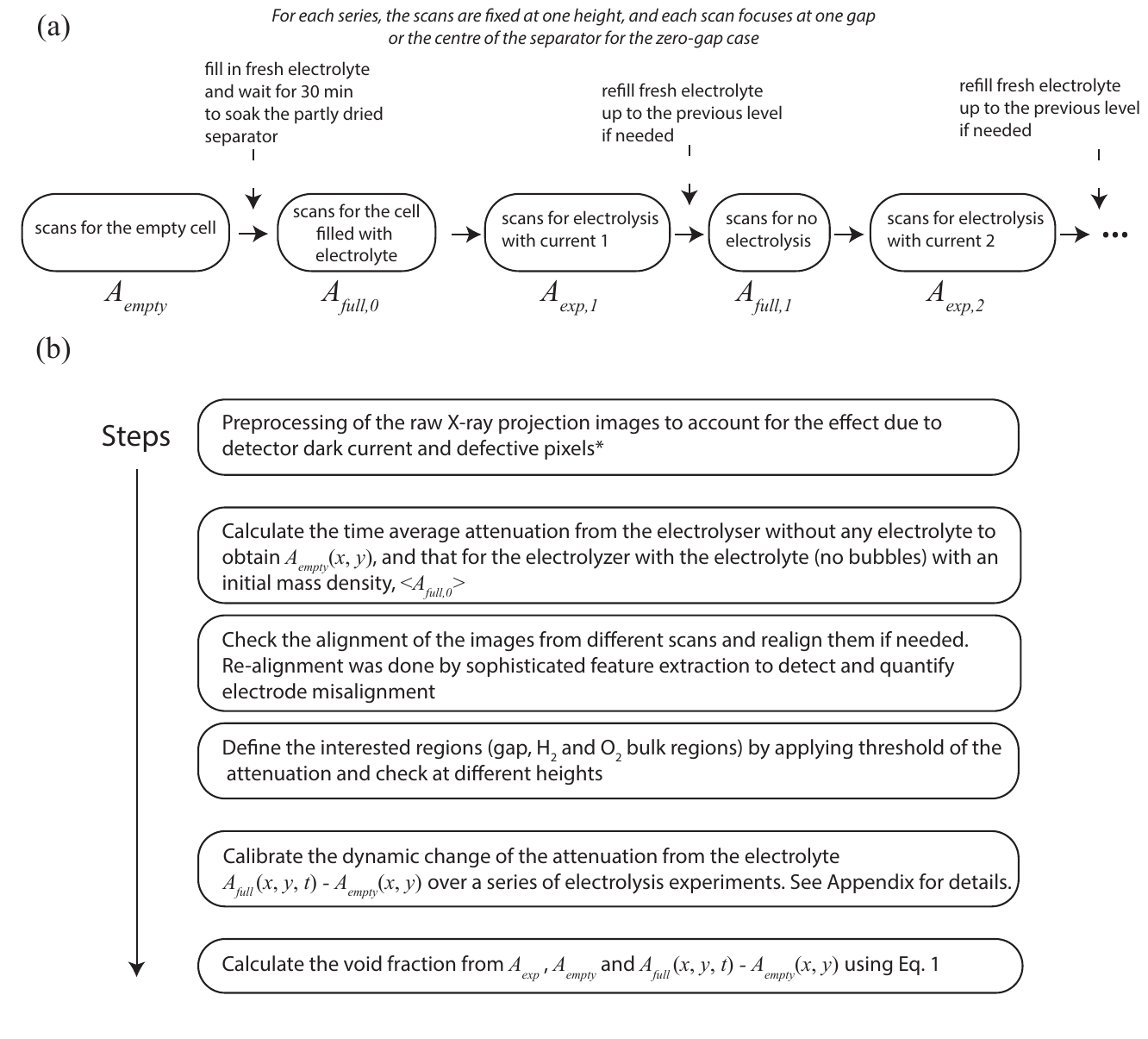}
    \caption{(a) Experimental procedure of a series of electrolysis experiments with the scans and actions performed. Below each step, the corresponding attenuation of the X-ray is obtained. (b) Procedures for obtaining the void fraction from the raw X-ray images. *For details, we refer to the first paragraph of Sec. 3.2 (except the tilting part) in \cite{Porombka2021}.}
    \label{fig:exp_procedure}
\end{figure*}

For the description of the X-ray image analysis to obtain the void fraction and the experimental procedures, we refer Fig. \ref{fig:exp_procedure}. The raw X-ray projection images are preprocessed in such way that the detector's dark current signal firstly is subtracted and then the signals of defective pixels are replaced by average signals from neighbouring non-defective pixels (see Sec. 3.2 in Ref. \cite{Porombka2021}). Basically, the detectors signals linearly represent the local intensity $I(x,y)$ of the impinging X-ray beam in the detector plane with $(x,y)$ representing the 2D location of the pixels in the detector plane.

\begin{figure}
    \centering
    \includegraphics[width=0.9\columnwidth]{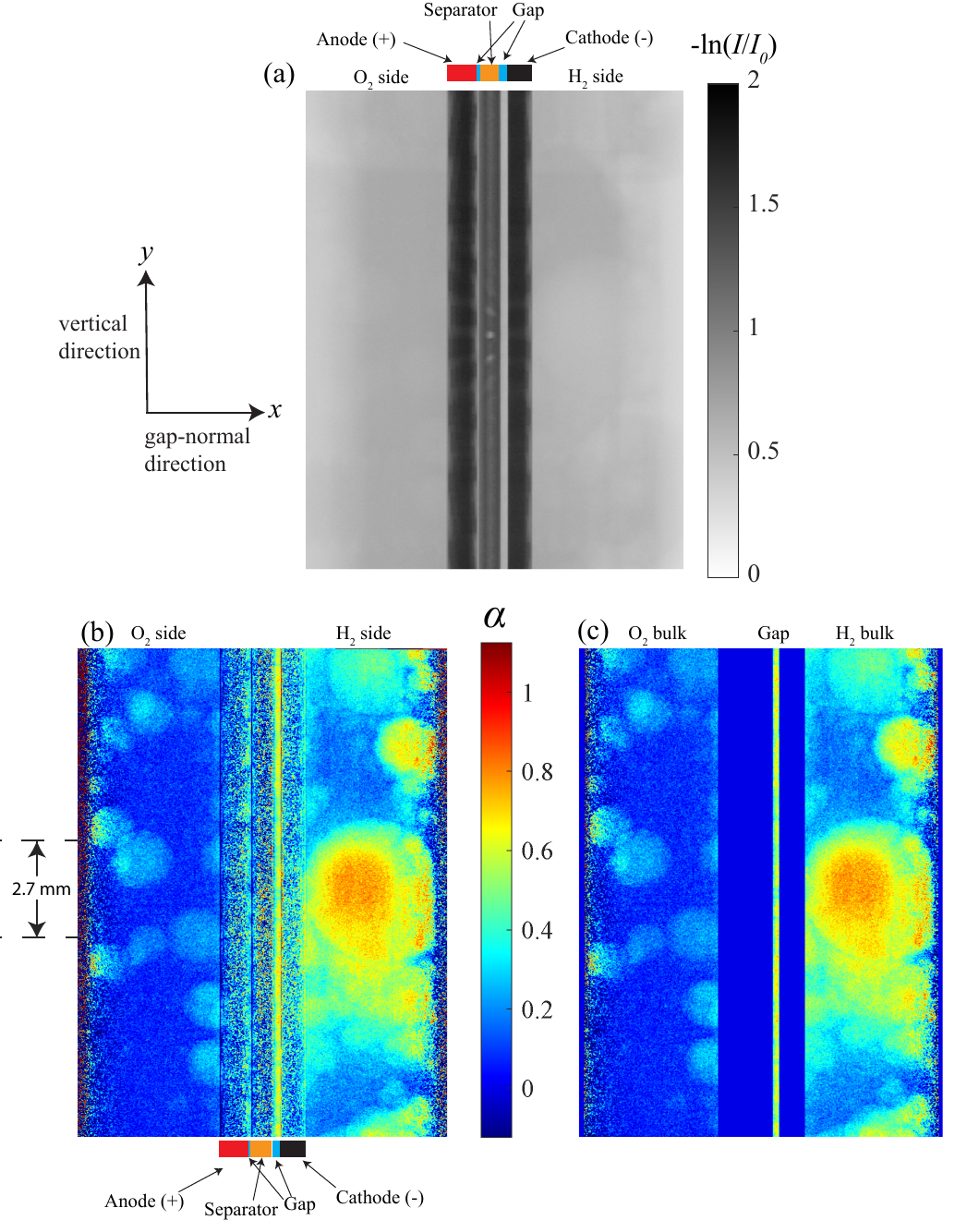}
    \caption{(a) Snapshot of the attenuation $-\ln(I/I_0)$ from radiography for the current density $j = \unit{0.16}{A.cm^{-2}}$, gap size $l_d = \unit{200}{\mu m}$, KOH concentration 24 wt\%, centering at height $H = \unit{60}{mm}$, and focusing at the H$_2$ gap (i.e. a side view). (b) The corresponding instantaneous void fraction $\alpha$ (Eq. \ref{eq:void_frac_definition}) for the attenuation image without applying a mask. (c) The corresponding instantaneous void fraction $\alpha$ for the attenuation image after applying a mask. }
    \label{fig:method_radioscopy_snapshot_example}
\end{figure}

We define the attenuation by $A = - \ln(I/I_0)$, where $I_0$ is the intensity of an X-ray beam emitted from the source and $I$ is the measured intensity of the X-ray beam after passing through the material between the source and the detector. The coordinate system $(x,y)$ is specified in Fig. \ref{fig:method_radioscopy_snapshot_example} which is parallel to the detector plate. Denote $t$ as the time coordinate. We define the attenuations $A_{empty}(x,y)$, $A_{full}(x,y,t)$, and $A_{exp}(x,y,t)$ as the attenuation measured when the electrolyzer has no electrolyte (except in the diaphragm) which is stationary, when the electrolyzer is full of electrolyte (no current), and during an electrolysis experiment, respectively. By Beer-Lambert's law, $A$ increases linearly with the thickness of a material an X-ray beam has passed through. This implies the liquid volume fraction $\alpha_L = (A_{exp} - A_{empty} )/(A_{full} - A_{empty})$, and the void fraction
\begin{equation}\label{eq:void_frac_definition}
    \alpha(x,y,t) = 1 - \alpha_L(x,y,t) = 1-\frac{A_{exp}(x,y,t) - A_{empty}(x,y)}{A_{full}(x,y,t) - A_{empty}(x,y)}.
\end{equation}

For each series of electrolysis experiments (see Fig. \ref{fig:exp_procedure}a for the whole experimental procedure), the cell's vertical position in the X-ray beam was fixed, and each scan focused at the H$_2$/O$_2$ gap or at the center of the separator for the zero-gap case (see Fig. \ref{fig:setup}f for the angles rotated for different gap sizes and gaps). The main experimental procedures were as follows: (1) scanning the empty cell to obtain $A_{empty}(x,y)$ averaged over the measurement time; (2) scanning of the cell filled with electrolyte ($A_{full,i}(x,y,t)$), where $i$ denotes the $i$-th electrolysis experiment and $i=0$ denotes the initial `full' case; and scanning of the cell at a current density $j$.

Note that in between two electrolysis experiments, if we observed any liquid level drop, fresh electrolyte was refilled to maintain a consistent initial level for all electrolysis experiments for each series of measurements. In addition, the chemical reactions resulted in a net consumption of H$_2$O and a time-dependent density of the electrolyte. This implies that $A_{full}-A_{empty}$ is changing over time. Thus, we established a time dependent scaling factor to account for this dynamic change of electrolyte density in the determination of void fraction (see \ref{sec:append_calc_void} for details). For the whole procedure on obtaining the void fraction from X-ray raw images, we refer to Fig. \ref{fig:exp_procedure}b.

Because of the 2D nature of X-ray radioscopy, the gap region is defined as the thin rectangular zone between the diaphragm on one side and the electrode and its perforations on the other side. The void fraction in the gap region is then the 2D projection of the 3D void fraction in this region, such that each point of the 2D projection represents an average over a line perpendicular to the 2D image.

An example of the attenuation image of an electrolysis experiment and the corresponding void fraction distribution is shown in Fig. \ref{fig:method_radioscopy_snapshot_example}, obtained by first obtaining the attenuation profile (Fig. \ref{fig:method_radioscopy_snapshot_example}a), then the void fraction profile without any masks (Fig. \ref{fig:method_radioscopy_snapshot_example}b), and finally with the masks (Fig. \ref{fig:method_radioscopy_snapshot_example}c). There are artefacts near the electrodes (nonphysical red patches and lines), which are resulted from insufficient correction of the misalignment of the electrodes between reference and electrolysis measurement due to electrode motion. We refer to \ref{sec:append_calc_void} for the details of the masking and the determination of the void fraction in the gap region. 


\section{Results}

The parameter space is spanned by different gap sizes $l_{gap}$, vertical cell positions in the X-ray beam $H$, and current densities $j$. For each pair of $l_{gap}$ and $H$, we performed electrolysis experiments for $j =  \{0.01, 0.16, 0.27, 0.48, 0.54 \}$ A/cm$^{2}$. To investigate the height dependence, we fixed $l_{gap} = \unit{200}{\mu m}$ and varied $H = \{28,44,60,76\}$ mm, while keeping in mind the total height of the active electrode is $\unit{94}{mm}$. To investigate the gap size dependence and unless otherwise specified, we fixed $H = \unit{60}{mm}$ and varied $l_{gap} = \{0,100,200,300\}$ $\mu$m. The initial KOH concentration for each series of electrolysis measurements (fixed $l_{gap}$ and $H$ but varying $j$) is $24$ wt\%. For the statistical notation, unless otherwise specified, we use a bracket $\langle \alpha \rangle$ as the spatial average and an upper bar $\overline{\alpha}$ as the time average.

\subsection{The overall structure of the bubbly flows, typical void fraction dynamics and typical IV curves}\label{sec:overall_structure}

\begin{figure}
    \centering
    \includegraphics[width=1\columnwidth]{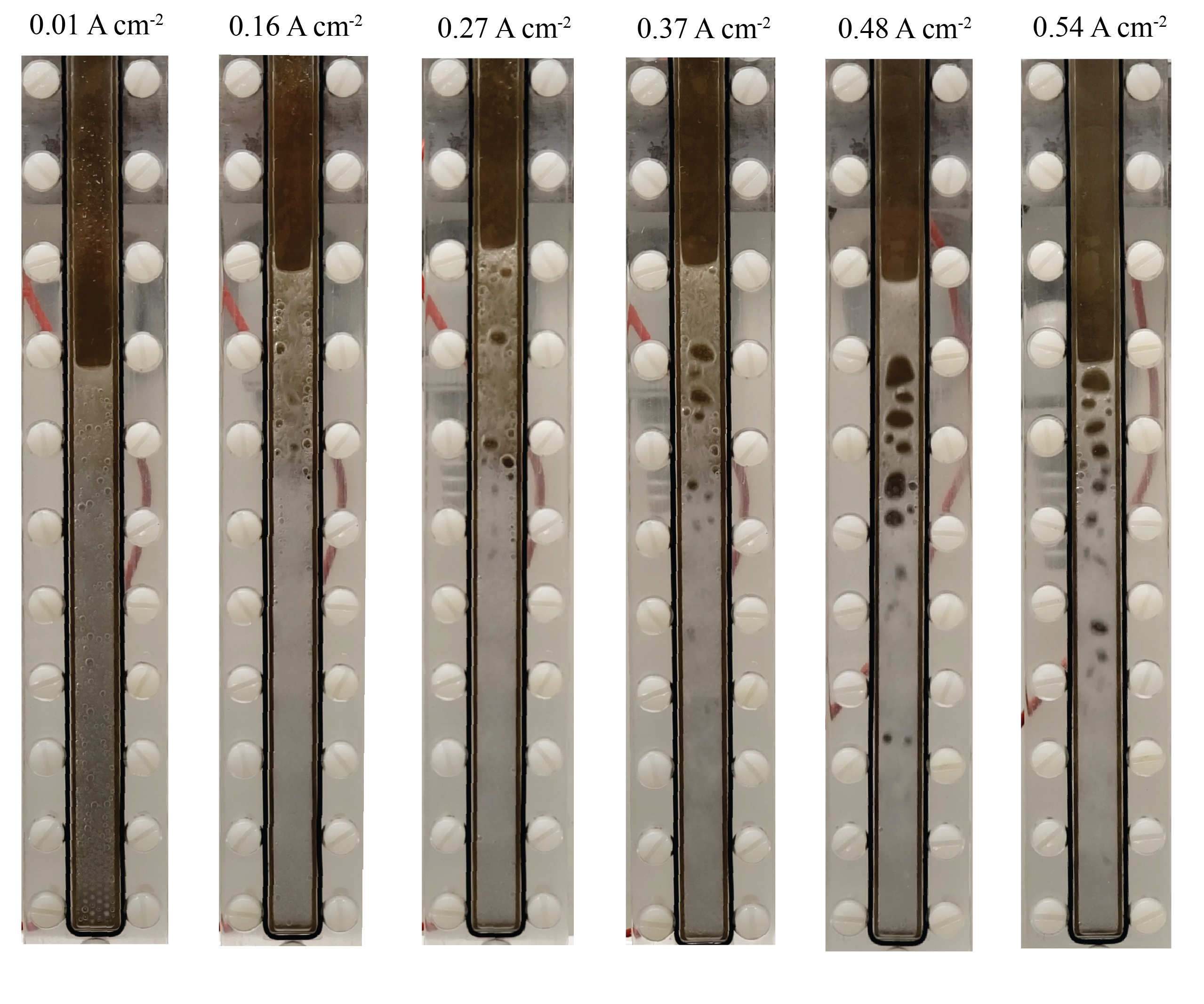}
    \caption{Snapshots of the front view of the bubbly flows of the electrolyzer for different current densities (increasing to the right). The snapshots here are only for reference with no indication of which chamber is referring to (the four cases from the left have the same chamber facing the camera and the two cases from the right have the same). We refer to the Supplementary material for the movies. }
    \label{fig:snapshot_bubbly_flows_camera}
\end{figure}

\begin{figure}
    \centering
    \includegraphics[width=1\columnwidth]{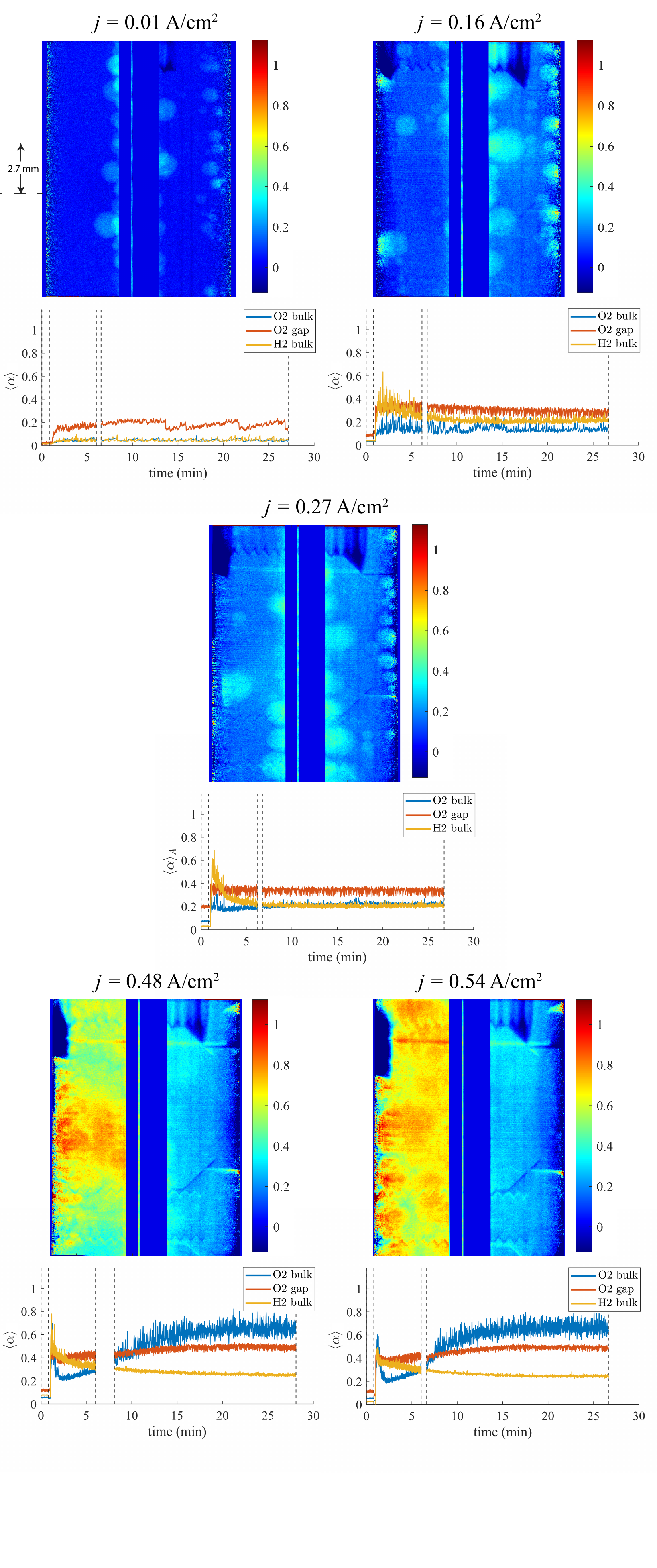}
    \caption{An illustration of the spatially averaged void fraction $\langle \alpha \rangle$ versus time $t$ for different current densities $j$ with the corresponding snapshots at around $t = \unit{17}{min}$. The radioscopy for this set of measurements focuses at the oxygen (O2) gap with the gap size $\unit{200}{\mu m}$ (i.e. side view). The spatial average of $\alpha$ is taken over $\pm \unit{1.4}{mm}$ from the mid-height of the image, as indicated next to the $j = \unit{0.01}{A/cm^2}$ image. The shape of a screw is visible at the top and the bottom of the radioscopy images (see Sec. \ref{sec:void_frac_and_exp_procedure} for the explanation). }
    \label{fig:overallVoidTimeExample_CTsnapshots}
\end{figure}

The overall structures of the bubbly flows are qualitatively examined. The snapshots of the bubbly flows of the whole electrolyzer from current densities ranging from $ j =\unit{0.01}{\ampere /\centi\meter^{2}}$ to $j = \unit{0.54}{\ampere/ \centi\meter^{2}}$ are shown in Fig. \ref{fig:snapshot_bubbly_flows_camera}. As expected, the bubbly flow gets denser as more gas is produced with increasing current density. One interesting observation is that coalescence clearly occur in our experiments, especially at the upper part of the cell. The coalescence results in big bubbles, or even plugs at current density higher than $\unit{0.48}{A/cm^2}$. This might be due to the geometrical dimensions of the cell. Moreover, the void fraction seems to increase over the cell height, which provides an expectation to the X-ray measurement that will be presented in the next section. We note that the structure of the bubbly flows are different from the hydrogen and oxygen chambers and even unsteady for high current densities (see Fig. \ref{fig:overallVoidTimeExample_CTsnapshots}). We refer to the Supplementary material for the movies.

Fig. \ref{fig:overallVoidTimeExample_CTsnapshots} shows examples of an instantaneous void fraction profile and the time trajectory of the spatially-averaged void fraction $\langle \alpha \rangle$ over $\pm \unit{1.4}{mm}$ at the middle of the scan at different current densities. The movies of such radioscopy scan are provided in the Supplementary material, too. We distinguish between the gap region (the radiography focuses at the O$_2$ gap in Fig. \ref{fig:overallVoidTimeExample_CTsnapshots}), the hydrogen (H$_2$) bulk region and the oxygen (O$_2$) bulk region. For $\alpha$ measured in the gap region, denoted as $\alpha_{gap}$, we emphasize that it is obtained by inferring the projections of the regions not only between the separator and the electrode material but also the region which faces the holes of the electrode (see Fig. \ref{fig:setup}a), meaning that we include the gas also facing the holes but also facing the electrode material (i.e. not the true void fraction between the electrode material and the separator). The spatial average $\langle \alpha_{gap} \rangle$ has an absolute error of $\pm 5 \%$ due to the possible movement of the electrolyzer during the scan. The explanation of this absolute error and the calculation of $\alpha_{gap}$ and its statistics are given in the \ref{app:void_frac_gap}.

When the current density increases, we expect more gas to be produced per unit time because of higher rate of the electrochemical reactions. 
The typical snapshots and time series of the void fraction in Fig. \ref{fig:overallVoidTimeExample_CTsnapshots} show that the void fraction indeed increases with increasing current density. Growing bubbles are observed at the cell walls. This can be due to the PMMA surface wetting properties (a water droplet contact angle of $\unit{68}{\degree}$ \cite{MA2007Polymer}) which may promote bubbles sticking at the wall. Another observation from the void fraction time series is that, except for the lowest current density $j = \unit{0.01}{\ampere. \centi\meter^{-2}}$, there is a build up of the void fraction after turning on the current, which is characterised by a sudden rise to a peak and a subsequent decay. 

The measured void fraction at the gap reaches to a steady value within 10 min at all investigated current densities. However this is not the case for the void fractions at the bulk. At $ j \leq \unit{0.27}{A/cm^2}$, both the H$_2$ and O$_2$ bulk void fractions become steady within 10 min, and their ratio is close to the stoichiometrics of the hydrogen evolution reaction (HER) and the oxygen evolution reaction (OER). At higher current densities ($j = \unit{0.48}{A/cm^2}$ and $\unit{0.54}{A/cm^2}$), there is a crossing behavior of these two bulk void fractions, i.e. the O$_2$ bulk fraction keeps increasing and H$_2$ bulk fraction keeps decreasing, resulting in a much higher void fraction in the O$_2$ chamber than that in the H$_2$ chamber. This behavior is interesting and will be further elaborated in Sec. \ref{sec:cross_flow}.


\subsection{Effect of gap size, current density and height dependence on void fraction}
\label{sec:gap_size_effect_to_bubble}

\begin{figure}
    \centering
    \includegraphics[width=1\columnwidth]{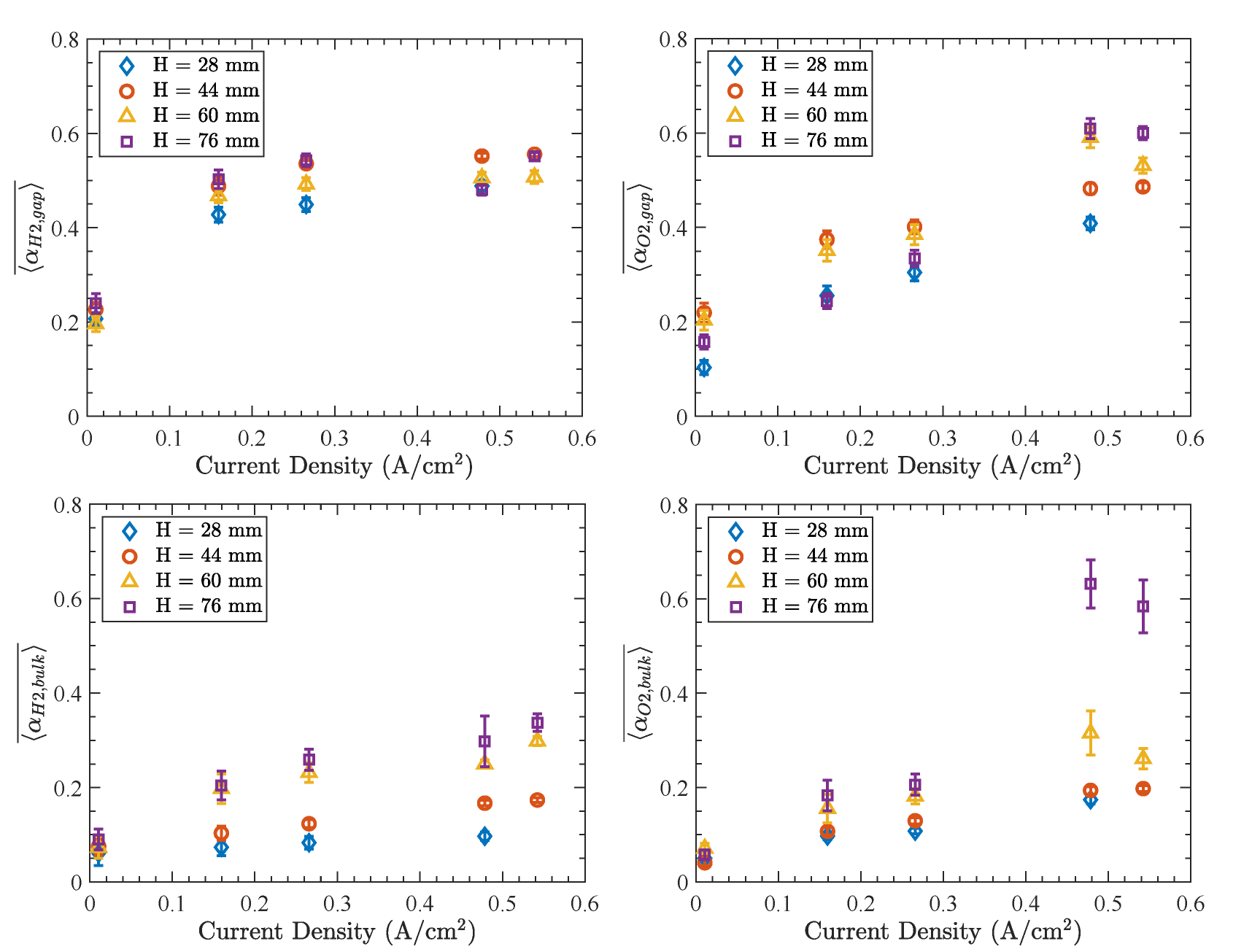}
    \caption{Mean void fractions at H$_2$ gap ($\overline{\langle\alpha_{H2,gap}\rangle}$), O$_2$ gap ($\overline{\langle\alpha_{O2,gap}\rangle}$),  H$_2$ bulk region ($\overline{\langle\alpha_{H2,bulk}\rangle}$) and O$_2$ bulk region ($\overline{\langle\alpha_{O2,bulk}\rangle}$) vs. current density at different heights of the cell with a gap size $l_{gap} = \unit{200}{\mu m}$. Multiple realisations for the same gap and same current densities are averaged. The error bars represents the propagated errors from individual standard deviations of the spatial averaged void fraction over time. For $\overline{\langle\alpha_{H2,gap}\rangle}$ and $\overline{\langle\alpha_{O2,gap}\rangle}$, the absolute error is $\pm 5 \%$ due to the possible movement of the electrolyzer during the scan (see \ref{app:void_frac_gap} for explanation).}
    \label{fig:alpha_vs_j_differentHeights}
\end{figure}

This section characterizes quantitatively the dependence of the void fraction on the gap size, the cell height, and the current density. Before reporting the results, we would like to make few remarks on the data analysis for proper interpretation of the results.

First, each scan is conducted with a focus at either H$_2$ gap, O$_2$ gap or the center of the membrane (for zero-gap case). \(\alpha_{H2,bulk}\) and \(\alpha_{O2,bulk}\) can be measured by focusing on any of these regions, and the results between focusing at the two gaps are similar in an electrolysis experiment (not shown here). Therefore, in our analysis, we use the \(\alpha_{H2,bulk}\) and \(\alpha_{O2,bulk}\) values obtained by focusing at the H\(_2\) gap or the center of the membrane (for zero-gap case). Time averaging is performed for the void fraction data at the last 10 min of the electrolysis experiment, where statistical equilibrium is achieved for $j \leq \unit{0.27}{A/cm^2}$, except for $j = \unit{0.01}{A/cm^2}$ in which the average is taken already from at 5 min after turning on the current. We note that for $j \geq \unit{0.48}{A/cm^2}$, statistical equilibrium is \textit{not} always achieved. As mentioned in the previous section, for $\alpha_{H2,gap}$ and $\alpha_{O2,gap}$, the absolute error is $\pm 5 \%$ due to the misalignment of the electrolyzer during the scan (see \ref{app:void_frac_gap} for details).


Fig. \ref{fig:alpha_vs_j_differentHeights} shows the mean void fractions at different heights along the cell with a gap size of $\unit{200}{\mu m}$ for different current densities. Overall, the void fraction at the gaps is consistent along the cell height. In contrast, the void fractions at the bulk show a clear height dependency, namely the gas holdup increases along the height as bubbles flow upwards. Despite some uncertainties in the data acquired at high current densities (as discussed in Sec. \ref{sec:overall_structure}), all these observations agree with the expectations. With regard to the dependency on current density, the overall void fraction in the cell increases with increasing density except for the void fraction at the H$_2$ gap. $\alpha_{H2,gap}$ quickly reaches a plateau of around 0.5.

\begin{figure}
    \centering
    \includegraphics[width=1\columnwidth]{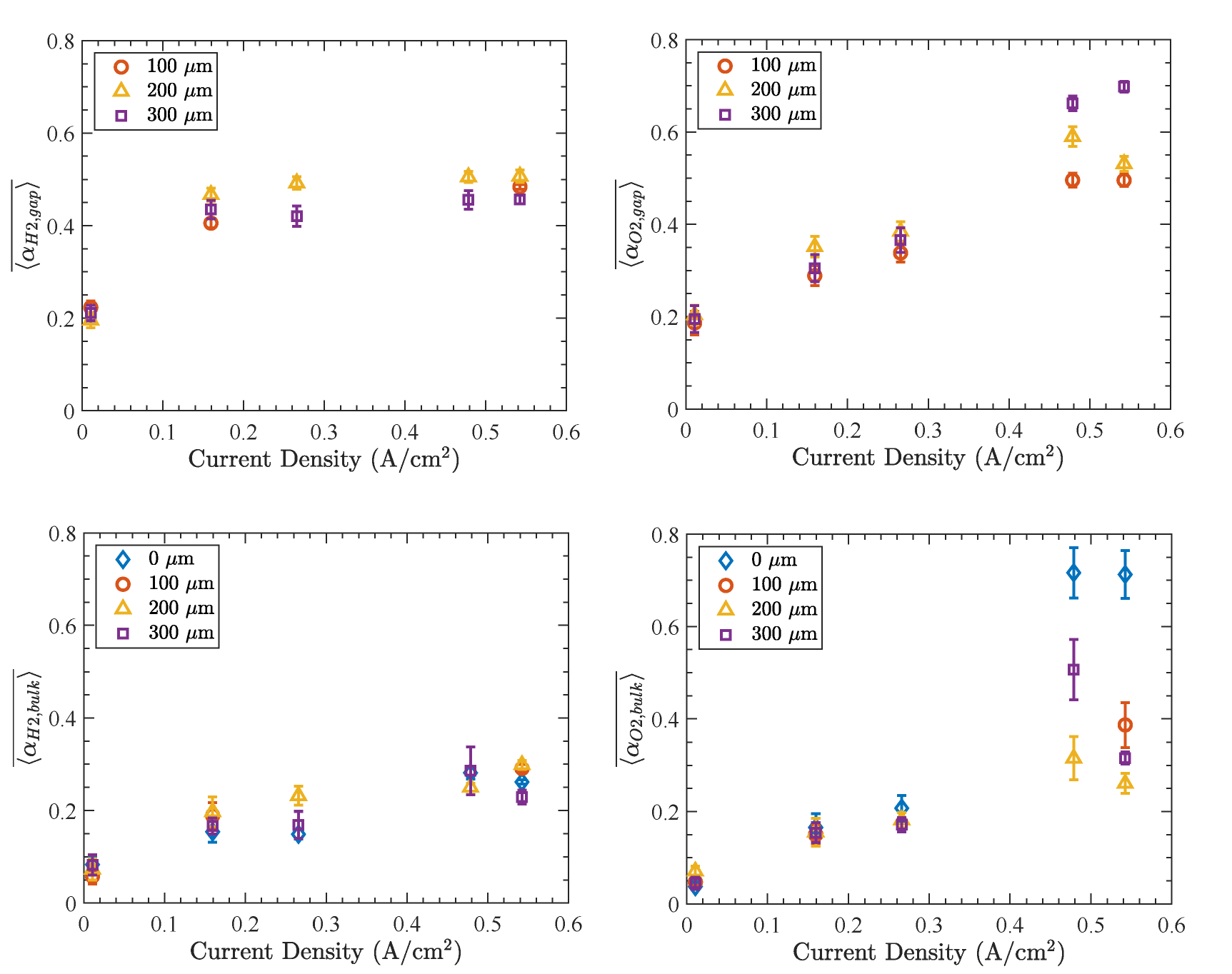}
    \caption{Mean void fractions at H$_2$ gap, O$_2$ gap, H$_2$ bulk region and O$_2$ bulk region (symbols same as Fig. \ref{fig:alpha_vs_j_differentHeights}) vs. current density at $H = \unit{60}{mm}$ of a cell with different gap configurations.}
\label{fig:alpha_vs_j_differentGaps}
\end{figure}

The mean void fraction $\overline{\langle \alpha \rangle}$ (time average of ${\langle \alpha \rangle}$) versus current densities for different gap sizes at a cell height $H = \unit{60}{mm} $ is shown in Fig. \ref{fig:alpha_vs_j_differentGaps}. Similar dependency on the current density can be seen in this figure as well. If excluding the data at the bulk at high current densities (due to uncertainties discussed earlier), we do not see a clear gap size dependency of the void fraction in the cell. 


\subsection{Effect of gap size and bubbles on cell voltage}\label{sec:gapSize_bub_effect_Ecell}

\begin{figure}
    \centering
    \includegraphics[width=1\columnwidth]{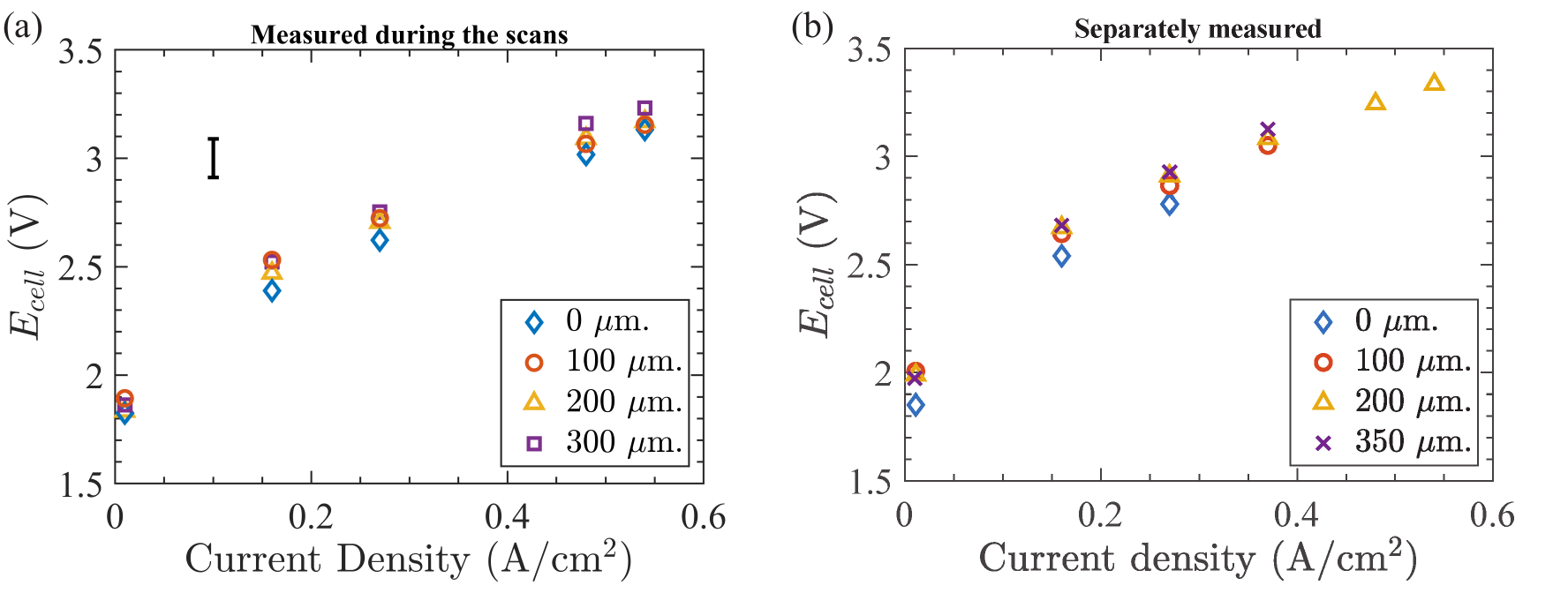}
    \caption{The mean cell voltage $E_{cell}$ vs. current density for different gap sizes for the experiments performed (a) during the CT and radiography scans, and (b) at a different occasion but similar temperature for each current density. For (a), each symbol is an average of different realizations for a given current density and given gap size, and the temperature for each realization is not precisely known, in contrast to (b) which each data point is obtained only through one realisation and the temperature only depends on the current density but not the gap size. The error bar in (a) of $\unit{0.2}{V}$ denotes the spread of the mean cell voltages for different realisations. For each cell voltage measurement, the standard deviation of the voltage time series is of $\mathcal{O}(\unit{0.01}{V})$ which is smaller than the marker size, thus not shown in the plot.}
    \label{fig:Ecell_vs_j_Dresden_TUe}
\end{figure}

The mean cell voltage $E_{cell}$ vs. current density for different gap sizes are shown in Fig. \ref{fig:Ecell_vs_j_Dresden_TUe}. Fig. \ref{fig:Ecell_vs_j_Dresden_TUe}a show the results measured during the X-ray scanning experiments. It should be noted that the temperature per test might vary since there is no temperature control of the setup. The results in this plot are reported as an average of three experimental realizations under each operating combination of $j$ and $l_{gap}$. As shown by the error bars, there exist a relatively large standard deviation ($\unit{0.2}{V}$), which is mainly due to temperature differences among different realizations. Therefore, to make a valid comparison, we performed the cell voltage experiments separately at a different occasion without the radiography scans but now with similar temperature for each $j$ and different $l_{gap}$, as shown in Fig. \ref{fig:Ecell_vs_j_Dresden_TUe}b, in which each data point is obtained only through one realisation. For each cell voltage measurement, the standard deviation of the voltage time series is of $\mathcal{O}(\unit{0.01}{V})$, which is smaller than the marker size thus not shown in the plot. 

These results show that the zero gap case is the most efficient, although only slightly. This observation differs from the findings in Ref. \cite{Haverkort2021} where the best performance was found with a gap distance of \unit{200}{\mu m}, possibly via stimulating bubble escape from the gap. We assume that these different findings are attributed to the cell configuration. For the cell built in Ref. \cite{Haverkort2021}, bolts tightening is applied to electrode and membrane, whereas in our cell only bolts tightening the cell are used to seal the O-ring for leakage prevention. In our zero-gap case, membrane and electrodes are just placed directly next to each other without being squeezed by bolts. This leads to different compression of the cell, which has been shown in Ref. \cite{Phillips2016} to influence on the cell voltage.

From the CT scan, we verify that the gap (if existing) between the diaphragm and electrode for our zero-gap case is within the spatial resolution of $\unit{15}{\mu m}$. We cannot absolutely disregard the possibility that there can be still small amount of electrolyte in between the electrodes and the diaphragm below this spatial resolution even in the zero gap case and possible replenishment with the bulk electrolyte over time. 

Furthermore, we try to quantify the contribution of gas/bubbles to the cell resistance in our measurements. First, Ohm's law is applied to determine the voltage drop across the gap, which is given by
\begin{equation}\label{eq:ohms_law}
    |j| = \lambda \frac{E_{gap}}{l_{gap}},
\end{equation}
where $j$ is the current density, $\lambda$ is the effective conductivity of the bubbly electrolyte solution in the gap, $E_{gap}$ is the potential drop across the gap and $l_{gap}$ is the width of the gap. Both, H$_2$ gap and O$_2$ gap contribute to the total cell potential drop which is given by,
\begin{equation}
    E_{gap,\text{total}} = E_{gap,\text{H}_2} + E_{gap,\text{O}_2}.
\end{equation}

\begin{figure}
    \centering
    \includegraphics[width=0.5\columnwidth]{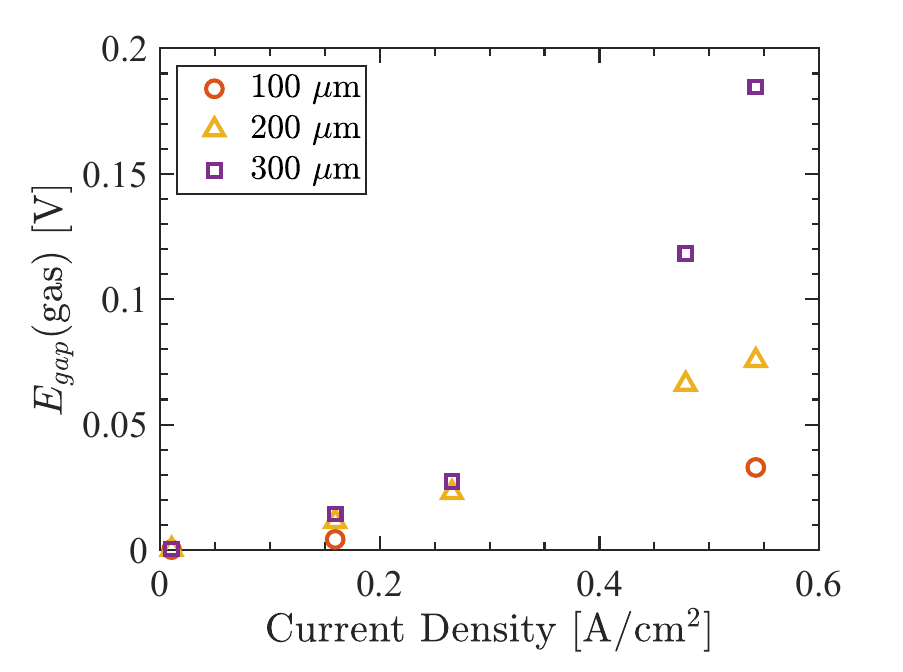}
    \caption{Estimated cell potential drop due to presence of gas/bubbles at the gap for different gap configurations at different current densities (see Eq. \ref{eq:Ecell_bub_drop} for the definition).}
    \label{fig:Egap_Bruggeman_total}
\end{figure}

The estimation of the effective conductivity in Eq. \ref{eq:ohms_law} follows the Bruggeman's model $\lambda_i = \kappa(1- \overline{\langle \alpha_{i,gap} \rangle} )^{3/2}$, where $\kappa$ is the conductivity of the electrolyte \cite{Tjaden2016}. By applying $\overline{\langle \alpha_{i,gap} \rangle}$ with a zero value or the measured value as shown in Fig. \ref{fig:alpha_vs_j_differentGaps}, we obtain $E_{gap} (0)$ and $E_{gap}(\overline{\langle \alpha_{i,gap} \rangle})$, which considers the pure electrolyte and the bubbly electrolyte, respectively. The difference corresponds to the contribution of bubbles to the cell potential drop
\begin{align}\label{eq:Ecell_bub_drop}
    E_{gap,bub}\equiv E_{gap}(\overline{\langle \alpha_{i,gap} \rangle})- E_{gap} (0) 
\end{align}
The results are shown in Fig. \ref{fig:Egap_Bruggeman_total}. As shown in the figure, $E_{gap,bub}$ increases with the current density, as a result of the increasing void fraction. Moreover, $E_{gap,bub}$ also increases with increasing $l_{gap}$ following Eq. \ref{eq:ohms_law} as $\overline{\langle \alpha_{i,gap} \rangle}$ is nearly insensitive to the gap size (except for $\overline{\langle \alpha_{O2,gap} \rangle}$ when $j \geq \unit{0.48}{A/cm^2}$, see Fig. \ref{fig:Egap_Bruggeman_total}). Therefore, the effect of bubbles to the cell voltage becomes more prominent at higher current densities and larger gap sizes. The largest contribution (0.2 V) as shown in Fig. \ref{fig:Egap_Bruggeman_total} corresponds to 6\% of the total cell voltage shown in Fig. \ref{fig:Ecell_vs_j_Dresden_TUe}.



\begin{table*}
\begin{center}
  \begin{tabular}{|c|c|c|c|c|}
    \hline
      electrode type & liquid contact angle* & porosity & thickness \unit{}{(\milli\meter)}  \\[3pt]
      \hline
Nickel plate & \unit{11}{\degree} to \unit{65}{\degree} \cite{HORSTHEMKE1985277} & 0.4 & 0.5 \\
\hline
Nickel foil-PMMA  & \unit{67.8}{\degree} (for PMMA)\cite{MA2007Polymer} & $0.08 \pm 0.01^{**}$ & $0.5+\mathcal{O}(0.1)$  \\
\hline
  \end{tabular}
  \caption{Properties of the electrodes used in this study. * refers to the side facing the bulk region at the chamber. **The uncertainty is due to the uncertainty from the hole size (the diameter is between 0.1 mm to 0.12 mm).}
  \label{tab:electrode_properties}
  \end{center}
\end{table*}

\subsection{Bubbly flow at high current density and cross-separator liquid transport}
\label{sec:cross_flow}

\begin{figure}
    \centering
    \includegraphics[width=0.9\columnwidth]{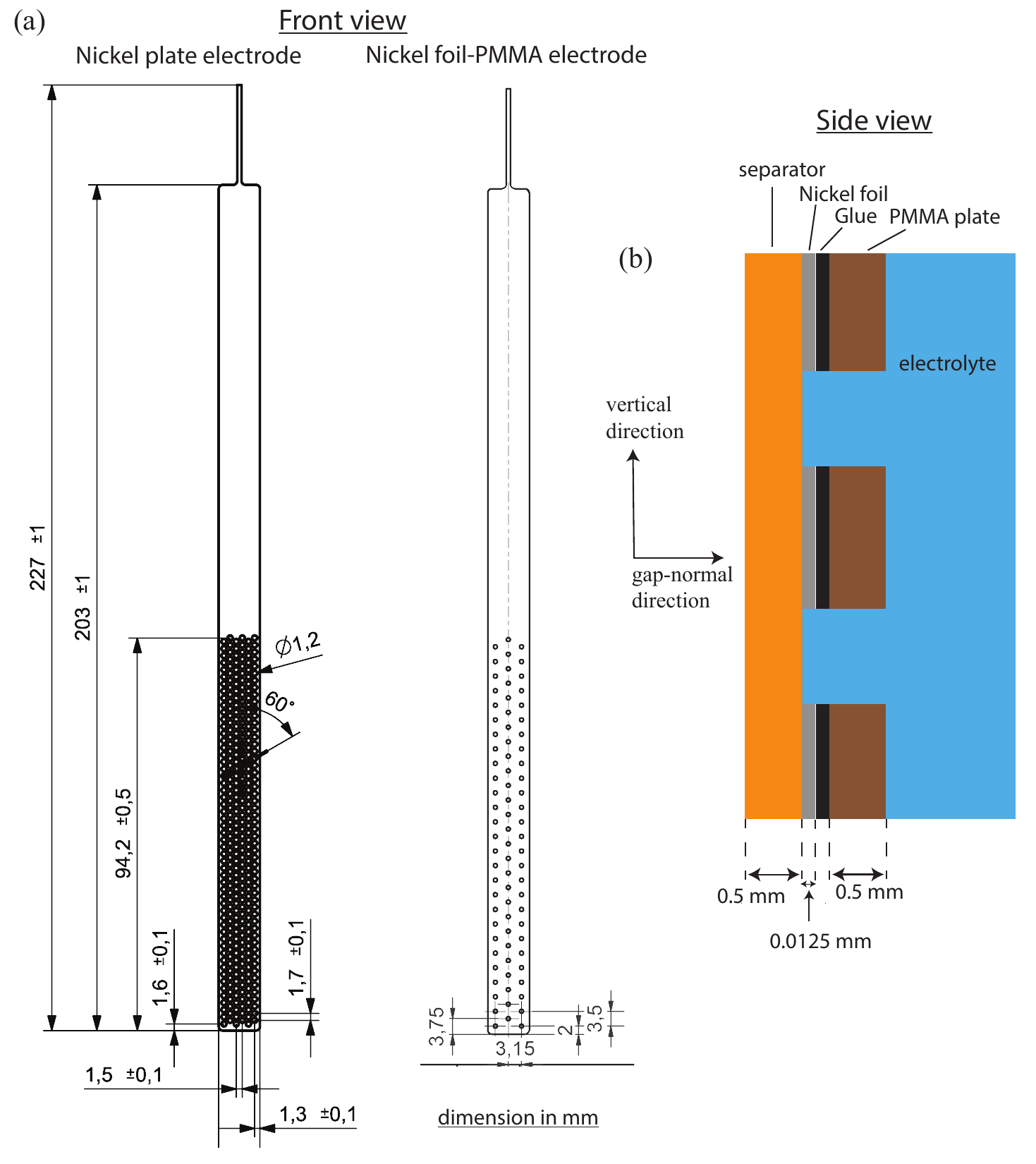}
    \caption{(a) Dimensions and illustration for porous nickel plate and foil-PMMA electrode at the front view. The distance between the holes for the foil-PMMA electrode is double than that in the plate electrode. (b) Configuration of the assembly of the foil-PMMA electrode at the right-half of the zero gap cell. The assembly for the whole cell is approximately symmetric about the center of the separator.}
    \label{fig:foil_plate_drawing_configuration}
\end{figure}

\begin{figure}
    \centering
    \includegraphics[width=0.9\columnwidth]{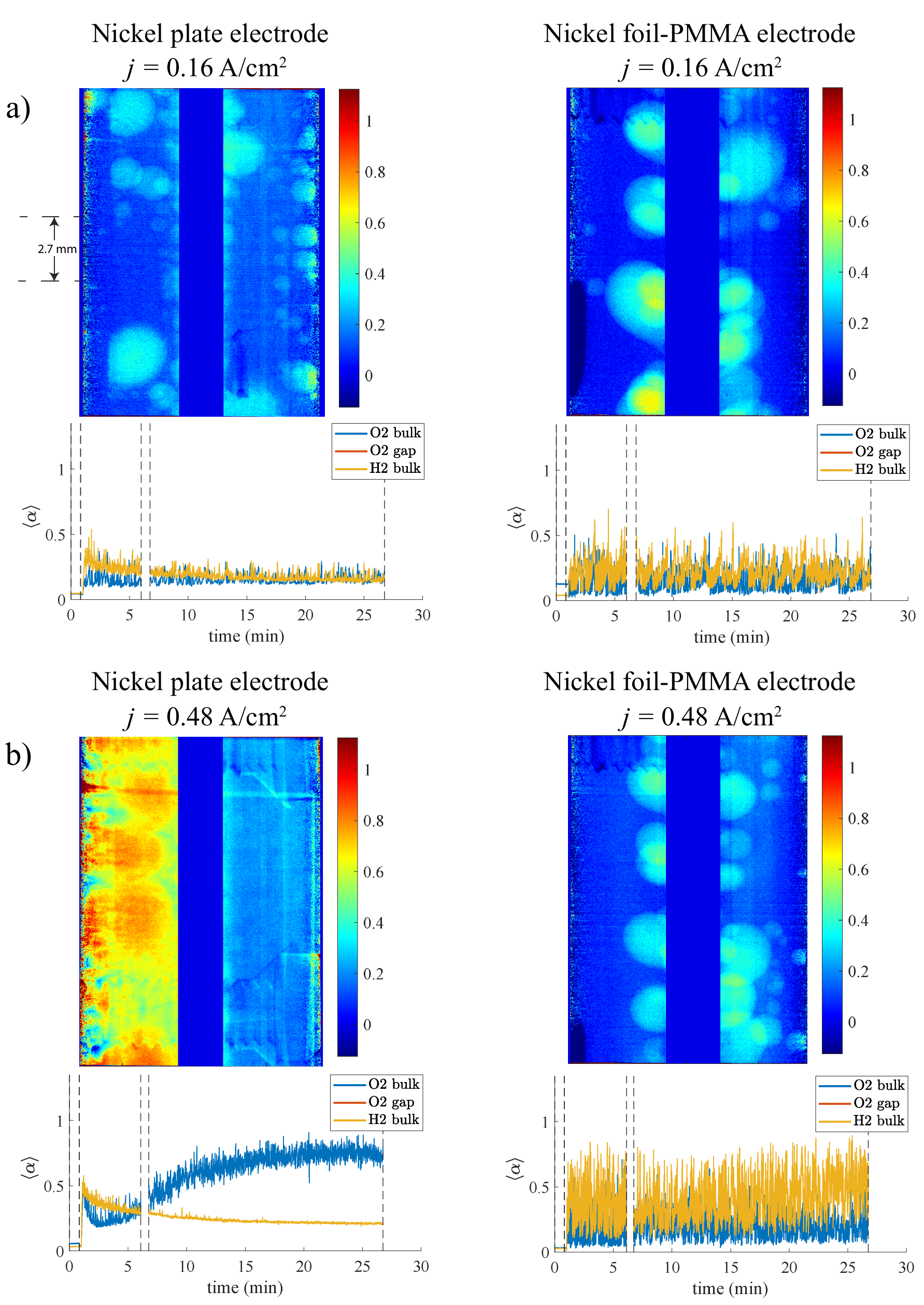}
    \caption{Spatial-averaged void fraction versus time in one set of measurements with two different current densities applied with the radiography snapshot at around 20 min for nickel plate electrode (left) and nickel foil-PMMA electrode (right). Both, the scanning heights are centred at $H = \unit{60}{mm}$. Their initial KOH concentrations are both 24 wt\% and the gap sizes are zero.}
    \label{fig:foi_plate_compare}
\end{figure}

A crossing behavior of the bulk void fraction has been observed at relatively high current densities as shown in Fig. \ref{fig:overallVoidTimeExample_CTsnapshots}. The void fraction at the O$_2$ compartment (both at the gap and the bulk) increases over time while that at the H$_2$ compartment decreases over time, resulting in a much higher void fraction at the O$_2$ compartment than at the H$_2$ compartment. However, this behavior does not occur when we change the electrodes from porous nickel plate to nickel-foil supported by a PMMA plate (see the dimensions in Fig. \ref{fig:foil_plate_drawing_configuration}). Fig. \ref{fig:foi_plate_compare} compares the void fraction profiles measured from experiments using these two types of electrodes. The comparison gives two key findings: 1) bigger bubbles are formed with foil-PMMA electrodes, resulting in larger fluctuations of the mean void fraction, 2) the cross behavior at high current densities with porous Nickel plate electrode is absent in the case of the foil-PMMA electrode. We refer to the movies of the radiography in the Supplementary material.


We believe these observations can be explained by difference of physical properties of these two electrode types, as listed in in Table \ref{tab:electrode_properties}. First nickle plate/foil and PMMA have different wettability (note that in foil-PMMA configuration it is the nickel foil facing the membrane and PMMA facing the bulk side). As for the contact angle shown in Table \ref{tab:electrode_properties}, the PMMA surface is less hydrophilic as compared to the nickel surface. This wettability difference well explains the bigger bubbles formed at the foil-PMMA electrode surface as seen in Fig. \ref{fig:foi_plate_compare}. Apart from that, the pressure between the diaphragm and the electrode for the foil case is higher than the electrode case due to the practicality of the assembly (stronger tightening force needed to keep the cell liquid tight).

Secondly, the foil-PMMA electrode has a much lower porosity ($\sim 0.1$) as compared to the nickel pate electrode (0.5). We think this less porous structure reduces the electrolyte and dissolved gas crossover phenomenon, thus prevents the crossing of the void fraction curves as observed in Fig. \ref{fig:overallVoidTimeExample_CTsnapshots}. To confirm, we try to quantify the liquid crossover by examining the liquid level changes at both compartments. Denote the change of liquid level at the O$_2$ side after an electrolysis experiment as $\Delta h_{O2}$ and that at the H$_2$ side as $\Delta h_{H2}$. The initial liquid levels of the two compartments are nearly the same. The stoichiometry for the chemical reaction of water electrolysis shows that H$_2$ side consumes double the amount of water produced at the O$_2$ side. If there is no transport between two compartments, $\Delta h_{O2} >0$ and $\Delta h_{H2}<0$, which implies always a positive value of $\Delta h_{O2} -\Delta h_{H2}$. In contrast, a negative $\Delta h_{O2} -\Delta h_{H2}$ implies a net transport of liquid/gas from the O$_2$ side to the H$_2$ side, even overcoming the hydrostatic pressure difference. 

\begin{figure}
    \centering
    \includegraphics[width=0.7\columnwidth]{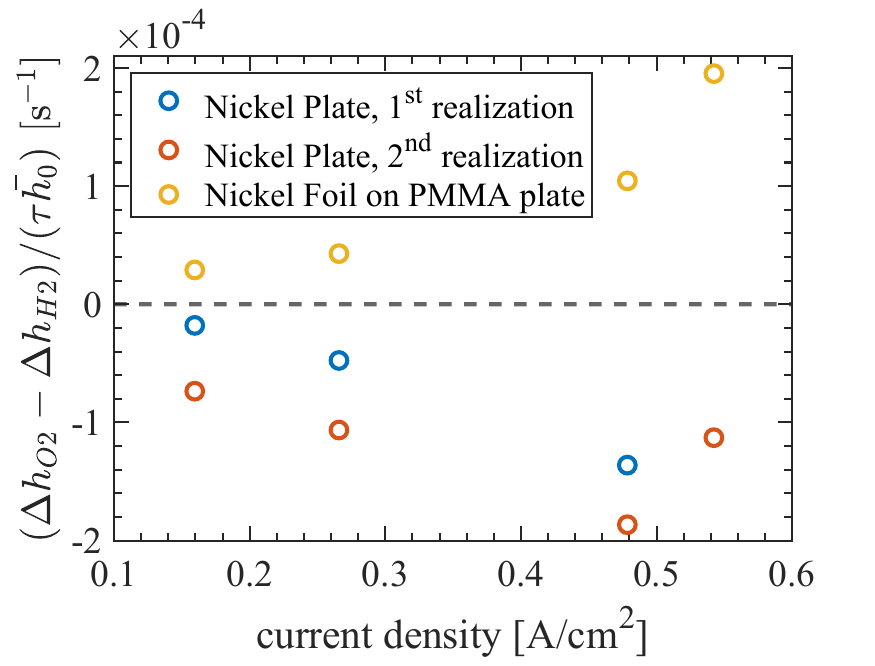}
    \caption{Change of liquid level at O$_2$ side $\Delta h_{O2}$ minus change of liquid level at H$_2$ side $\Delta h_{H2}$ after an electrolysis experiment, normalized by the average initial liquid level before any electrolysis experiment $\overline{h_{0}}$ and the duration of the electrolysis experiment $\tau$. These three sets of experiments are all zero gap with KOH 24 wt\%. We note that the liquid level change results might also account for (1) small lost of liquid via splashing out at the top of the cell in particular at high current densities and (2) bubbles remained in the cell before and after the electrolysis experiments. These effect are similar for the same current density, meaning that the comparison between the foil and the plate electrode cases is valid.}
    \label{fig:h_diff_dt_foil_plate}
\end{figure}

Fig. \ref{fig:h_diff_dt_foil_plate} shows the results of the liquid level change analysis in a zero-gap cell with nickel porous plate or foil-PMMA electrodes at different current densities. For the porous plate electrode case, $\Delta h_{O2} -\Delta h_{H2}<0$ is observed (even for other gap sizes in nearly all experiments, though not shown here). This implies a significant liquid transport from the anodic compartment (O$_2$ side) to the cathodic compartment (H$_2$ side). The same phenomenon was observed and discussed in previous literature \cite{Haverkort2020PRA,Haverkort2020JPowerAdv,gennes2004capillarity}. They attribute this crossover to the electro-osmotic flow, which is driven by the movement of positive ions (K$^+$) at the electric-double layer due to the presence of an electric potential difference, and becomes significant when the potential difference is large and the size of the channels is small with high porosity. As the current density increases with potential differences, the electro-osmotic flow becomes stronger with increasing current densities. This is indeed the case as seen in Fig. \ref{fig:h_diff_dt_foil_plate} that the crossover rate is enhanced with increasing current density for an porous nickel plate electrode. 

In contrast, for the much less porous foil-PMMA electrode, $\Delta h_{O2} -\Delta h_{H2}$ is found always positive in the figure. This implies a limited liquid transport between two compartments. In addition to the lower porosity, bigger bubbles grow (longer residence time) at the holes on the (PMMA) surface, as explained earlier, which might also explain this transport limit.

To summarize, the analysis of liquid level change in electrolysis experiments clearly reveals an enhanced liquid crossover from the anodic compartment to the cathodic compartment in our (small-size) cell using a relatively large porous electrode at higher current densities. This might provide a convincing explanation to the cross behaviour of the two bulk gas fractions from experiments with the nickel plate electrode, while is was not observed from experiments with the foil-PMMA electrode, as seen in Fig. \ref{fig:overallVoidTimeExample_CTsnapshots} and Fig. \ref{fig:foi_plate_compare}.

\subsubsection{Effect of the cross-separator liquid flow}

The liquid crossover for the porous plate electrode leads to excess liquid loses/depletion at the anodic compartment, while a constant volumetric flux of gas production is retained. As a result, the measured void fraction from X-ray scanning increases in time, which becomes apparent at higher current densities. In the cathodic compartment, however, the crossover leads to a liquid gain with a constant depletion due to HER, thus resulting in a slightly decreasing void fraction in time. As the crossover is suppressed using the foil-PMMA electrode, the void fraction measured in the two compartments stays rather steady, though with a slow increase in the void fraction in the cathodic compartment due to the faster liquid depletion there.

As discussed in Ref. \cite{Haverkort2020JPowerAdv}, the presence of such cross-separator flow can lead to undesirable crossover of dissolved oxygen, and this can be suppressed by hydrostatic pressure difference corresponding to approximately 0.1 m of liquid level difference \cite{Haverkort2020PRA}. Note that we estimate a net velocity of $\unit{1.5}{\mu m /s} $ normal to the separator due to this liquid crossover, which is close to the electro-osmotic velocity of $\unit{1}{\mu m /s}$ reported in \cite{Haverkort2020JPowerAdv}, although that was for different molar concentration ($c = \unit{0.1}{M}$) and current density ($j = \unit{0.126}{A.cm^{-2}}$). The detailed estimation is shown in Appendix \ref{app:crossover}.

\section{Conclusions and outlook}

To the best knowledge of the authors, this work for the first time reveals the void fraction distribution in a zero/narrow gap alkaline water electrolyzer using X-ray radiography. With these measurement, we clear up some speculations/hypothesis in the open literature regarding the effect of gas bubbles/film on the cell performance as concluded below. The experiments have been carried out using an in-house built $\unit{8.1}{mm}\times \unit{7.8}{mm} \times \unit{198}{mm}$ cell with adjustable gap sizes ranging from $\unit{0}{\mu m}$ to $\unit{300}{\mu m}$ using KOH concentration 24 wt\% at different operating current densities varied between $\unit{0.01}{A/cm^2}$and $\unit{0.54}{A/cm^2}$.

First of all, the measured gas fraction is reported individually at anodic compartment (O$_2$ side), cathodic compartment (H$_2$ side), anodic gap, and cathodic gap. The overall void fraction increases with increasing current density. The void fraction at the gaps are found to be rather stable/consistent along the cell height regardless of the gap size. Our measurements do not show evidence on bubble trapping or gas film formation in the gap as speculated in literature.  
The void fraction at the bulk is found to increase along the cell height, however it was nearly insensitive to the gap size. Interestingly, a crossover behavior of the bulk void fractions, i.e. $\overline{\langle\alpha_{O2}\rangle}$ increasing in time and $\overline{\langle\alpha_{H2}\rangle}$ decreasing in time, is observed with a significantly higher void fraction at the anodic compartment in our experiments at higher current densities. This is a consequence of liquid crossover/transport between the two compartments with an enhancement by high porosity of the electrodes, which is evidenced by the liquid level change comparing cases with electrodes of different porosities.

Furthermore, our voltage measurements imply that a zero gap configuration (within $\unit{15}{\mu m}$ resolution) is still most efficient compared to finite small gaps up to $\unit{300}{\mu m}$. Applying the Bruggeman model using the measured gas fraction, the cell potential drop due to the presence of bubbles/gas at the gap is estimated to be maximum 6\% of the total cell voltage at $j = \unit{0.54}{A. cm^{-2}}$ up to a gap size $ l_{gap} = \unit{300}{\mu m}$. This finding does not agree with the suggestion e.g. in Ref. \cite{Haverkort2021} of introducing a gap of $\unit{200}{\mu m}$ to improve the cell efficiency by allowing easier escape of the bubbles. However, we do stress that the different ways of the cell assembly in our work and in Ref. \cite{Haverkort2021} might result in different compression conditions, which can play an important role in the cell performance and are subject to further investigation.

Finally, liquid crossover from the O$_2$ to the H$_2$ side was observed like in Ref. \cite{Haverkort2020JPowerAdv}.  In the current study, we found that the crossover was suppressed when the electrode porosity was reduced and a thin film electrode attached to a PMMA surface with lower wettability was applied.

Last but not least, we have faced quite some challenges in X-ray measurement and data processing, e.g., strong artefacts bear to the electrodes because of huge scattering of the X-ray was at the electrode of a high mass density, and complicated correction required during data processing as a consequence of not completely refreshed electrolyte in the experiment series. Future study should address these issues for a refined experimental/measurement design. For example, to maintain the concentration of the electrolyte solution, one can connect the chambers with a pump and a reservoir of electrolyte. To reduce the scattering of X-rays at the electrode, one may employ a thinner and/or less dense electrode. However, finding such a material that is chemically resistant to the electrolyte (KOH) can be challenging, and a thinner electrode can induce higher ohmic heating.

\vspace{0.5cm}
\noindent
{\bf CRediT authorship contribution statement}

\noindent
On-Yu Dung:  Writing – original draft, Formal analysis, Visualization, Conceptualization, Investigation.
\\
Stephan Boden: Writing - review \& editing, Data curation, Software, Formal analysis, Visualization, Resources, Investigation, Methodology.
\\
Albertus W. Vreman: Writing - review \& editing, Conceptualization, Supervision.
\\
Niels G. Deen: Writing - review \& editing, Supervision, Project administration.

\noindent
Markus Schubert: Writing - review \& editing, Project administration, Resources.

\noindent
Yali Tang: Writing - review \& editing, Conceptualization, Supervision, Fund acquisition, Project administration.

\vspace{0.5cm}
\noindent
{\bf Declaration of generative AI in scientific writing}

During the preparation of this work the author(s) used ChatGPT in order to correct the grammar of certain sentences. After using this tool/service, the author(s) reviewed and edited the content as needed and take(s) full responsibility for the content of the publication

\vspace{0.5cm}
\noindent
{\bf Acknowledgement}

This project is funded by NWO (ECCM KICKstart DE-NL program, KICH1.ED04.20.014) and co-funded by Nobian. Hans Lammers (Nobian) is acknowledged for suggesting foil-PMMA electrodes. Jan Timmers and Bart van Sassen are acknowledged for the design of the prototype and the improvement to the final design of the electrolyzer, respectively. We acknowledge the technician Hans van Griensven from the Power and Flow group for giving advices on the design, Henri Vliegen and Thijs van Mol for the mechanical support, and the Eindhoven Prototype Center (EPC) (Gerrit Fimerius, Luciano Oorthuizen, and Bram Kerkhoff) for building the components of the setup. We also acknowledge the technicians Nico Bley, André Flaisz, Jan Schäfer, Ronny Berger, and Uwe Sprewitz at HZDR for the technical support of the X-ray experiments at HZDR. Frank Barthel and Dominic Windisch are also thanked for their patience with us since we occupied and blocked the X-ray laboratory almost all the time. We thank Prof. Uwe Hampel for approval and the facilitation of this project at HZDR.

\appendix


\section{X-ray image processing and the dynamic modelling of the attenuation across the electrolyte}\label{sec:append_calc_void}

This section provides the details of the procedures of obtaining the void fraction from the X-ray raw images in Fig. \ref{fig:exp_procedure}. Firstly, the raw X-ray projection images are pre-processed to account for the effect due to detector dark current and defective pixels, and we refer the details to the first paragraph (except the tilting part) in section 3.2 in Ref. \cite{Porombka2021}. The masks for the gap and the bulk regions are obtained by making thresholds to the change of the attenuation along a horizontal height at the `empty' reference scan and check it for different height locations. 



\subsection{The void fraction in the gap region and its absolute error}\label{app:void_frac_gap}
The gap size is only 7 pixels ($\unit{100}{\mu m}$) to 20 pixels ($\unit{300}{\mu m}$) and the gap is close to the artefacts produced by beam-hardening near the electrodes. Therefore, we only consider the void fraction at the center of the gap region, assuming it is homogeneous at the gap, to deduce the void fraction. Thus we obtain the void fraction on a vertical line ($\alpha_{gap}(y,t)$), which represents the projection of a vertical plane parallel to the diaphragm and the electrode. From the time series of $\langle \alpha_{gap} \rangle_y(t)$ of certain electrolysis measurements, we found that $\langle \alpha_{gap} \rangle_y(t)$ sometimes jumps up or down for $5\%$. This is possibly due to the movement of the cell. Therefore, we take the absolute error of $\langle \alpha_{gap} \rangle_y$ as $\pm 5\%$.

\subsection{General strategy of the dynamic modelling of the attenuation across the electrolyte}

To calculate the void fraction, we employ Eq. \ref{eq:void_frac_definition} which requires the knowledge of dynamic behavior of the attenuation across the electrolyte, 
$A_{full}(x,y,t) - A_{empty}(x,y)$. The mass density of the electrolyte changes since overall water is being consumed for water electrolysis while the mass of KOH remains unchanged in the solution. In general, such change of density depends on spatial coordinate but we assume that there is perfect mixing at each chamber (H$_2$ or O$_2$ chamber) due to the mixing by bubbles. Then, we model the attenuation across the electrolyte by 
\begin{equation}\label{eq:scaling_factor_def}
    A_{full}(x,y,t) - A_{empty}(x,y) = k(t) ( A_{full,0}(x,y) - A_{empty}(x,y) ),
\end{equation}
where $k(t)$ is a scaling factor, modeled separately for the H$_2$ and the O$_2$ chamber, and $A_{full,0}(x,y)$ is the attenuation measured when the electrolyzer is full of electrolyte before any electrolysis experiments which is stationary.

We note that we cannot measure the attenuation across the electrolyte without bubbles during the electrolysis because bubbles were present everywhere in the cell. We arrived this conclusion by checking the attenuation measured during the measurement and found that almost all region has a lower attenuation than that without electrolysis, i.e. there were always some bubbles present in any region during the measurement. However, we did perform a scan over the electrolyte \textit{before} each electrolysis measurement (with the caution that we filled in fresh electrolyte on top which may not be mixed before the electrolysis starts, see Fig. \ref{fig:exp_procedure}a for the procedure). Therefore, we determine $k(t_{1,i})$ at the start time ($t_{1,i}$) and $k(t_{2,i})$ at the end time ($t_{2,i}$) of electrolysis experiment $i$ and use linear interpolation to obtain $k(t)$ in between these two times. 

\subsection{The mixing between fresh and electrolyte present in the electrolyzer}

To determine $k(t_{1,i})$ and $k(t_{2,i})$, we use the `full' scans performed before electrolysis experiments $i$ and $i+1$, respectively. 
First, we note that each `full' scan is performed at some distance below the liquid level, while the fresh electrolyte is added at above the view of the scan. We need to assume whether, at one extreme, the mixing between the fresh and the one originally in the cell occurs during the $i$-th `full' scan or, to the other extreme, this mixing happens after the electrolysis starts. If the latter is true, then the attenuation measured just before the measurement starts in fact represents the attenuation at the end of the previous electrolysis experiment but not the initial electrolyte attenuation in the upcoming experiment. We checked a typical series of electrolysis measurement and found that there is only a few percent of the attenuation difference between the case of assuming the mixing happens already when the fresh one is filled into the cell and the case of assuming the mixing occurring only when the electrolysis starts to occur. Due to the comparatively long time scale of mixing due to pure diffusion, we assume the aforementioned mixing occurs only after the electrolysis starts. 

\subsection{Definition of the scaling factors before and after an electrolysis experiment}

From the experimental procedure shown in Fig. \ref{fig:exp_procedure}a, we obtain the attenuation of the `full' case $A_{full,i}$, where $i = 0,1,2,...$, which can then be used for inferring the attenuation across the electrolyte before an electrolysis measurement by calculating the difference between the `full' and `empty' attenuation. Because of the assumption in the previous subsection,  $k(t_{2,i})$ should be related to $A_{full,i+1}(x,y)-A_{\empty}(x,y)$.

However, in practice, since the reference `full' scans for $i \geq 1$ were performed after the previous electrolysis measurement (and filled with fresh electrolyte to maintain the same level), there were still bubbles stuck during these scans. For a short-hand notation, we denote the time-averaged attenuation across the electrolyte during the $i$-th `full' scan as 
\begin{equation}
    B_i(x,y)= A_{full,i}(x,y) - A_{empty}(x,y). 
\end{equation} 
As we only wish to find the attenuation across the electrolyte without bubbles, we use the maximum value of $B_i$ at each vertical column at the bulk (since the presence of a bubble will lower the attenuation), denoted by $\max_y(B_i)$ which then depends only on the horizontal coordinate $x$. If we assume there is negligible vertical dependence of the mass density of the electrolyte, $\max_y 
 (B_i)$ estimates the attenuation across the electrolyte. For $i =0$, the `full' scan has no bubbles and therefore the average over height $\langle B_0 \rangle_y$  represents the attenuation across the electrolyte itself, which also depends on $x$. The overall \textit{relative change} of $B_i$ across the electrolyte obtained from the $i$-th `full' scan is used to obtain $k(t_{2,i})$: 
\begin{equation}
    k(t_{2,i}) = \left\langle\dfrac{\max_y(B_i)}{\langle B_0 \rangle_y}\right\rangle_x,
\end{equation}
for $i = 1,2,...$, where $\langle \cdot \rangle$ is a spatial average over the $x$-direction. This is a constant, i.e. it does not depend on $x$ and $y$. If $B_i$ and $B_0$ do not depend on $x$ and $y$, then the definition above implies $k(t_{2,i})=B_i/B_0$, which is equivalent to $B_i = k(t_{2,i})B_0$. The latter is in fact equation (A.1) evaluated at time $t_{2,i}$. 

To infer $k(t_{1,i+1})$ from $k(t_{2,i})$, we took into account the mixing between the fresh and the electrolyte left in the electrolyzer after measurement $i$. Recall that we assume the mixing only occurs just after the electrolysis experiment starts. 


We modelled the attenuation across the KOH electrolyte for different values of the mass densities ($\rho$) using the X-ray spectra with source spectra \cite{hammersberg1998measurements} and X-ray attenuation coefficients from literature \cite{berger2010xcom}. The following correlation was obtained:
\begin{equation}
K(\rho) \equiv \frac{\langle \langle B \rangle_y \rangle_x}{\langle \langle B_0 \rangle_y \rangle_x}
\end{equation}
where $B(x,y)=A_{full}(x,y)-A_{empty}(x,y)$.  

Using the equation $K(\rho_{2,i})=k(t_{2,i})$, we first convert $k(t_{2,i})$ to the density $\rho_{2,i}$ of the electrolyte. The liquid levels before and after adding the fresh electrolyte are known and the density of the fresh electrolyte is also known. Assuming the total volume is unchanged before and after mixing, the density $\rho_{1,i+1}$ of the electrolyte after the mixing can then be approximated. Finally, 
$k(t_{1,i+1})$ is then obtained using 
\begin{equation}
k(t_{1,i+1})=K(\rho_{1,i+1})
\end{equation}
Thus, all $k(t_{1,i})$ and $k(t_{2,i})$ are determined, except the last one,  $k(t_{2,N})$, where $N$ is the number of electrolysis experiments in the series.  For $k(t_{2,N})$, it is obtained by extrapolating a linear fit of $k(t_{2,i})$ versus the accumulated electric charge applied after the $i$-th electrolysis experiment.

\section{Implementing the Bruggeman model to calculate the contribution to the cell voltage from the void at the gap}\label{sec:append_Egap}

By assuming spherical bubbles, the Bruggeman model assumes the effective conductivity as $\lambda = \kappa(1-\overline{\langle\alpha_{gap}\rangle})^{3/2}$, where $\kappa$ is the electrolyte conductivity and $\alpha$ is the void fraction at the gap \cite{Tjaden2016}. To estimate $\kappa$, the corresponding molarity and temperature of the electrolyte needs to be known \cite{OBrien2005handbook}. The molarity of the electrolyte is inferred from the X-ray attenuation across the electrolyte before the electrolysis occurs (see \ref{sec:append_calc_void} for the details). However, as mentioned before, due to the prevention of displacement of the setup from an intrusive temperature measurement which can lead to misalignment for the scan, the temperature was not measured. Therefore, to estimate $\kappa$, we assume the corresponding temperature $\unit{26}{\degreecelsius}, \unit{31}{\degreecelsius},\unit{29}{\degreecelsius},\unit{34}{\degreecelsius},$ and $\unit{32}{\degreecelsius}$ for the current densities $\unit{0.01}{A.cm^{-2}},\unit{0.16}{A.cm^{-2}},\unit{0.27}{A.cm^{-2}},\unit{0.48}{A.cm^{-2}}$, and $\unit{0.54}{A.cm^{-2}}$ respectively. Such rough estimation of the temperature are obtained from the experiments at a different occasion but at the same place where the CT scan was made, though the electrolysis duration time were not always the same as those performed during the CT scan. 


\section{The estimation of the liquid crossover flow speed}\label{app:crossover}

To investigate the potential effect of the cross-separator transport, we estimate the transport velocity as follows. With the presence of transport from the O$_2$ side to the H$_2$ side, by continuity, at the O$_2$ side there is an induced flow to be sucked to the diaphragm, contributing to a downward flow from the upper half of the cell since the upper-half the electrodes are insulated by a insulator strip. We first consider the theoretical production and consumption of water at each chamber due to chemical reaction, denoted as $P_{i}$, where $i =$ H$_2$ or O$_2$. After that, we estimate the liquid level change per unit time by the observed liquid level change at each chamber divided by the electrolysis duration, which gives us the rate of the net change of the volume of solution at each chamber, denoted as $\dot{V}_{i}$, where $i =$ H$_2$ or O$_2$. The liquid transport from one chamber to another $T_{k \rightarrow i}$ required to drive such net change is estimated by $T_{k \rightarrow i} = \dot{V}_{i} - P_{i}$, where $i,k =$ H$_2$ or O$_2$ and $i\neq k$. We estimate $\dot{V}_{i} = (\Delta h_i/\tau) \times S_i$, where $\tau$ is the duration of the electrolysis experiment and $S_i$ is the horizontal area of the chamber $i$. Accordingly, for the net velocity normal to the separator due to the transport for $j = \unit{0.54}{A.cm^{-2}}$, $u_{sep} = T_{k \rightarrow i} / S_{sep}  \approx \unit{1.5}{\mu m . s^{-1}}$, where $S_{sep}$ is the area of the diaphragm. The velocity is comparable to the electro-osmotic velocity of $\unit{1}{\mu m s^{-1}}$ found in Ref.  \cite{Haverkort2020JPowerAdv}, although that was for a different molar concentration ($c = \unit{0.1}{M}$) and current density ($j = \unit{0.126}{A.cm^{-2}}$).




\bibliographystyle{elsarticle-num} 
 \bibliography{zero_gap_bib}





\end{document}